\RequirePackage{lineno}
\documentclass[prb,twocolumn,aps,amsmath,amssymb]{revtex4}
\usepackage{graphicx}
\usepackage{bm}
\usepackage{ulem}
\begin{document}

\title{A new route to enhance the ferromagnetic transition temperature in diluted 
magnetic semiconductors}

\author{Kalpataru Pradhan$^1$, Subrat K Das$^2$}

\affiliation{$^1$CMP Division, Saha Institute of Nuclear Physics, HBNI, Kolkata 700064, India\\
  $^2$SKCG Autonomous College, Paralakhemundi, Odisha 761200, India}

\begin{abstract}
We investigate the magnetic and the transport properties of diluted magnetic 
semiconductors using a spin-fermion Monte-Carlo method on a 3D lattice in the 
intermediate coupling regime. The ferromagnetic transition temperature $T_c$ 
shows an optimization behavior, first increases and then decreases, with 
respect to the absolute carrier density $p_{abs}$ for a given magnetic 
impurity concentration $x$, as seen in the experiment. Our calculations 
also show an insulator-metal-insulator transition across the optimum $p_{abs}$ 
where the $T_c$ is maximum. Remarkably, the optimum $p_{abs}$ values lie in a 
narrow range around 0.11 for all $x$ values and the ferromagnetic $T_c$ 
increases with $x$. We explain our results using the polaron percolation 
mechanism and outline a new route to enhance the ferromagnetic transition 
temperature in experiments.
\noindent

\end{abstract}
\vspace{.25cm}

\maketitle

Diluted magnetic semiconductors (DMSs) are materials of strong interest due 
to both, their novel ferromagnetism and potentiality for future 
spintronics\cite{munekata,ohno1,dietl1,jungwirth,dietl2} . In particular, 
Ga$_{1-x}$Mn$_x$As\cite{ohno,dietl} with ferromagnetic transition temperature 
$T_c$ $\simeq$ 173 K in bulk\cite{jungwirth1}, $\simeq$ 191 K in films\cite{chen}, 
and even more ($\simeq$ 200 K)\cite{chen1} in nano wires has led intensive 
efforts to increase $T_c$ in view of possible technological applications.

It is widely accepted that Mn$^{2+}$ ion ($Mn_{Ga}$) replace Ga$^{3+}$ ion in 
Ga$_{1-x}$Mn$_x$As and thereby contributes a hole to the semiconductor valence 
band, which mediate the magnetic interaction between the localized spins.
However, point defects like $Mn$ interstitials ($Mn_{I}$) and As 
anti-sites\cite{yu,myers} significantly compensates the free hole density. 
In addition, $Mn_{I}$ are highly mobile and preferentially choose the 
interstitial positions adjacent to $Ga$ substituted $Mn$ ions ($Mn_{Ga}$), 
thus forming anti-ferromagnetic $Mn_{Ga}$-$Mn_{I}$ pairs\cite{blinowski} and 
consequently increases the $Mn$ inactive sites. So overall $Mn_{I}$ reduces 
the hole density and the effective $Mn$ concentration ($x_{eff}$) to 
$Mn_{Ga}$-2$Mn_{I}$ and $Mn_{Ga}$-$Mn_{I}$, respectively, hindering the 
higher ferromagnetic $T_c$ in DMSs. We define the carrier density 
$p$ = $p_{abs}$/$x$, where $x$ is the $Mn$ concentration and $p_{abs}$ is 
the absolute carrier density. The $p_{abs}$ in our language is similar 
to the hole density (holes per cm$^{3}$), generally reported in the 
experiments.

Most of the experimental studies have been devoted in search of room-temperature 
ferromagnetism using different methods. Post-growth annealing is one of the most 
extensively used technique which enhances the $T_c$ by reducing the $Mn_{I}$ 
concentration and in turn increasing the carrier concentration\cite{potashnik}.
It is important to note that all $Mn_{I}$ can not be 
removed from the sample\cite{edmonds}. Even after annealing the fraction of 
$Mn_I$ increases beyond 0.2 for $x$ $\sim$ 0.1, putting a limit to the $T_c$ 
which either saturates or decreases at larger $x$ \cite{potashnik1,dobrowolska}. 
Consequently, the system goes from insulating to metallic 
and then again to insulating phase with $x$\cite{matsukura}. 

Another route to enhance the $T_c$ is by p-type and n-type co-doping method that 
can tune the hole density. It is shown that with Be co-doping both $p$ and $T_c$ 
increase for low $x$ (=0.03), where as for $x$ $\gtrsim$ 0.05 $p$ saturates and 
$T_c$ decreases due to the increase in $Mn_I$ concentration\cite{lee,yu1}. In 
contrast, in Si co-doped samples $T_c$ systematically increases with $x$ even for 
$x$ $\gtrsim$ 0.07 [Ref.\citenum{cho,kim}]. It is found that $p$ has lower values 
in the Si co-doped samples for all values of $x$. For low $x$ ($\leq 0.08$) 
samples the $T_c$ is lower compared to the un-codoped ones\cite{cho,wang1}
and for higher $x$ the co-doped sample shows higher $T_c$. This enhancement in 
$T_c$ is attributed to the increase in the hole mobility in the impurity band.
In addition, for lower $x$ ($\leq 0.08$) and higher Si co-doped samples both 
$p$ and $T_c$ increase as Si start to replace $Mn_I$ sites.

It is believed that it is necessary to increase $Mn$ concentration to enhance 
the $T_c$\cite{jungwirth1}, but $T_c$ decreases beyond $x_{eff}$=0.07 
[Ref. \citenum{dobrowolska}]. So it is important to search a route in which 
both the $Mn$ concentration and the hole density can be altered using 
growth and post-growth technique. 
In this letter, we investigate this scenario and outline a procedure to enhance 
$T_c$ with $x$. We calculate the ferromagnetic transition temperature within 
a diluted Kondo lattice model in the intermediate coupling regime using a 
Monte-Carlo technique based on travelling cluster approximation\cite{tca-ref} 
on large size systems. The ferromagnetic $T_c$ shows an optimization behavior 
with $x$ and $p_{abs}$, and in the process system undergoes an 
insulator-metal-insulator transition. Our results qualitatively agree with 
the recent experiments. We find that optimum $p_{abs}$, where $T_c$ is maximum, 
lies around 0.11 for a wide range of $x$=0.1-0.35. For a fixed $p_{abs}$ 
ferromagnetic $T_c$ increases with $x$, which suggests a new pathway to 
achieve high temperature DMSs. 

We consider a particle-hole symmetric diluted Kondo lattice 
Hamiltonian\cite{alvarez,chattopadhyay,berciu} in 3D. The model is given by  

\[H=-
\sum_{{\langle ij \rangle} \sigma} t_{ij} c^{\dagger}_{i\sigma} c_{j\sigma}  
-{\dfrac{J}{2}}\sum_{k} {\bf S}_{k} .{\vec \sigma}_{k} -\mu\sum_i n_i, \]

\noindent where $t$ is the nearest neighbor hopping parameter, $\mu$ is the 
chemical potential, and $J$ ($> 0$) is the Hund's coupling between the localized 
impurity spins ${\bf S}_{k}$ and the itinerant electrons (${\vec \sigma}_{k}$) 
at random impurity sites ${k}$. We assume that the ${\bf S}_k$ to  be classical 
unit vectors.
We have considered simple cubic lattice with one atom per unit cell, where as 
GaAs is face centered cubic with four atoms per unit 
cell. So, roughly 25\% of $x$ in our case corresponds to 6.25\% that in real 
experiments. Magnetic moment clustering and hence the direct exchange interaction 
between the spins are neglected. 

In random systems like DMSs the theoretical calculations must adequate a larger
system size and also able to capture the effects of spatial fluctuations for a 
better estimation of any physical quantity such as $T_c$. Spin-fermion 
Monte-Carlo is an effective approach to take spatial fluctuations into 
account\cite{alvarez,pradhan}. An exact diagonalization scheme is applied to 
the itinerant carriers in the background of randomly located classical 
spins ${\bf S}_k$. In order to avoid size limitation we employ a Monte-Carlo 
technique based on travelling cluster approximation\cite{tca-ref,pradhan} to 
handle system size as large as $N$ = L$^3$ = 10$^3$. All physical quantities 
are averaged over ten different such random configurations. In our particle-hole 
symmetric model the magnetic and transport properties are presented in terms 
of the hole density. We set $t$ = 1 and all other parameters like $J$, 
temperature $T$ are scaled with $t$. 

Magnetic properties in DMSs are consequence of the competition between the 
carrier mediated ferromagnetic spin-spin interaction and the carrier 
localization. We start our calculation with a specific choice of $x$ = 0.25 
and $p$ = 0.5 which is a good starting point for simple cubic lattice.
For $J$ $\sim$ 0 there is no Mn-Mn interaction and for very large $J$ the carriers
are trapped in Mn sites. So ferromagnetism is suppressed in these two extreme limits
and the optimal $T_c$ lies in the intermediate range of $J$ as shown in Fig.1(a).
Carrier localization for higher $J$ leads to the developments of
an impurity like band in the density of states,  
$N(\omega)=\langle {1 \over N} \sum_\alpha \delta(\omega-\epsilon_\alpha) \rangle $ 
at relatively high temperature $T$=0.05, as shown in Fig.1(b). 
We estimate $T_{c}$ from the ferromagnetic structure 
factor $S(\textbf{0})$, where 
$S(\textbf{q})$ =${1 \over N}$ $\sum_{ij}$ 
$\bf {\bf S}_i\cdot {\bf S}_j$ e$^{i\bf{q} \cdot ({\bf r}_i-{\bf r}_j)}$. 
The ferromagnetic structure factor for $J$=5, $x$=0.25, and $p$=0.5 is plotted in 
Fig.1(c) and the inset shows that $S(\textbf{0})$ for L=8 and L=10 
are barely distinguishable at $T_c$. So we have considered L=8 for rest of 
our calculations.

\begin{figure}[t]
\centerline{
\includegraphics[width=9.0cm,height=7.7cm,angle=0,clip=true]{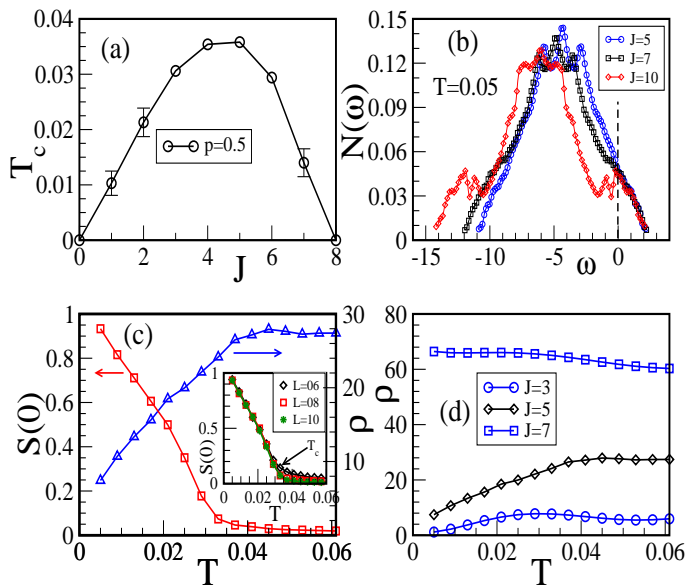}}
\vspace{.2cm}
\caption{Color online: (a) ferromagnetic transition temperature $T_{c}$ for 
different $J$ values at $p$=0.5; (b) density of states (DOS) at $T$=0.05 for 
$J$=5, 7, and 10 (Fermi energy is set at zero); (c) temperature dependence of 
the ferromagnetic structure factor $S(\textbf{0})$ and resistivity for 
$J$=5 and $p$=0.5. The inset shows $S(\textbf{0})$ for three different sizes 
L=6, 8, and 10. The arrow indicates the $T_c$; (d) temperature dependence of 
the resistivity for $J$=3, 5, and 7.
}
\end{figure}

We obtain the resistivity for different $J$ values by calculating the {\it dc} 
limit of the conductivity as determined by the Kubo-Greenwood 
formula\cite{mahan-book, cond-ref} as shown in Fig.1(d). At low temperature 
the system shows metallic behavior for small and intermediate $J$ values. As $J$ 
increases ($J = 7$) the system remains insulating in the whole temperature range 
due to carrier localization at impurity sites. Hereafter, we concentrate in the 
intermediate coupling regime ($J$ = 5) where $T_{c}$ is found to be maximum and 
relevant to DMSs. Temperature dependence of the resistivity and the ferromagnetic 
structure factor in Fig.1(c) show one-to-one correspondence between the onset of 
the ferromagnetism and the metalicity at $T_c$ = 0.037.

In carrier-mediated magnetic systems a minimum amount of carrier is essential to 
initiate the coupling between the magnetic spins, which depends on $J$ and $x$. 
On the other hand, for higher carrier density the magnetism is suppressed due to 
decrease in carrier mobility. The overall behavior of $T_c$ with $p$ is shown 
in Fig.2(a) for $J$=5 and $x$=0.25. The mobility picture is confirmed from the 
conductivity calculation, where $T_{c}$ and the conductivity (at the low 
temperature) are maximum at $p$=0.45 (see inset). In order to compare our result 
with the experiment we plot data from Ref. \citenum{dobrowolska} in Fig. 2(b) 
such that the impurity concentration $x$ lies in a narrow range from 
$0.025-0.035$ which we assume to be constant. Now if we match, the $T_c$ vs 
$p$ behavior in the experiment is very similar to our results. A metal-insulator 
transition with $p$ is also observed in the same experiment (not shown here) 
which we already illustrated the inset of Fig.2(a).

\begin{figure}[t]
\centerline{
\includegraphics[width=9.0cm,height=7.7cm,angle=0,clip=true]{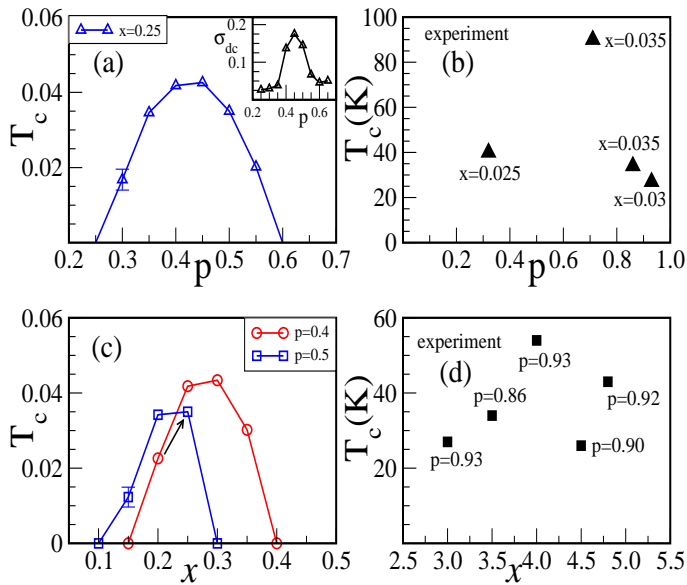}}
\vspace{.2cm}
\caption{
Color online: For $J$=5 (a) ferromagnetic transition temperature $T_c$ 
[inset: {\it dc} conductivity] for different $p$ values for $x$=0.25 
({\it dc} conductivity is calculated at T=0.005); (c) Ferromagnetic transition 
temperature $T_c$ for different $x$ values for $p$=0.4 and 0.5; (b) and (d) 
ferromagnetic transition temperatures $T_c$ taken from the experiment 
[Ref. \citenum{dobrowolska}]. For details please see the text. 
}
\end{figure}

The carrier mobility and hence the ferromagnetism can be tuned with $J$, $p$ 
or $x$ independently. The carrier-spin interaction $J$ is only operative at 
the impurity sites i.e., for fixed $J$ value the effective coupling strength of 
the system increases with impurity concentration  $x$. This is similar to the 
case of increasing $J$ with keeping $x$ fixed. So the variation of $T_{c}$ with 
$x$ for fixed $p$ values [Fig. 2(c)] can be understood from the $T_{c}$ 
dependence of $J$ as in Fig. 1(a). We plot the experimental data from 
Ref. \citenum{dobrowolska} in Fig. 2(d) such that the carrier density $p$ 
lies in a narrow range from $0.86-0.93$. We have neglected this small variation 
of $p$ for qualitative comparison with our calculations and found that the 
$T_c$ shows an optimization behavior with $x$, quite similar to our results. 
It is important to note that if we increase both $x$ and $p$ along the arrow 
shown in Fig. 2(c) the $T_C$ increases, which mimics the effect of post-growth 
annealing on $T_C$. 

\begin{figure}[t]
\centerline{
\includegraphics[width=9.0cm,height=8.7cm,angle=0,clip=true]{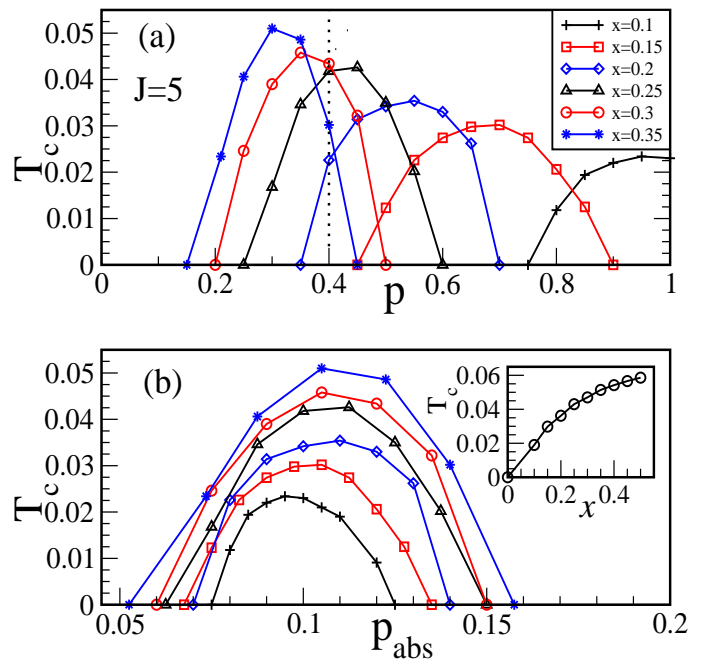}}
\vspace{.2cm}
\caption{
Color online: Ferromagnetic windows for different magnetic impurity concentrations 
$x$. (a) ferromagnetic windows in $p$ and (b) ferromagnetic windows in absolute 
carrier density, $p_{abs}$. Inset: variation of transition temperature with $x$ 
for fixed absolute carrier density, $p_{abs}$=0.11.
}
\end{figure}

\begin{figure*}[t] 
\centerline{
\includegraphics[width=18.0cm,height=5.0cm,angle=0,clip=true]{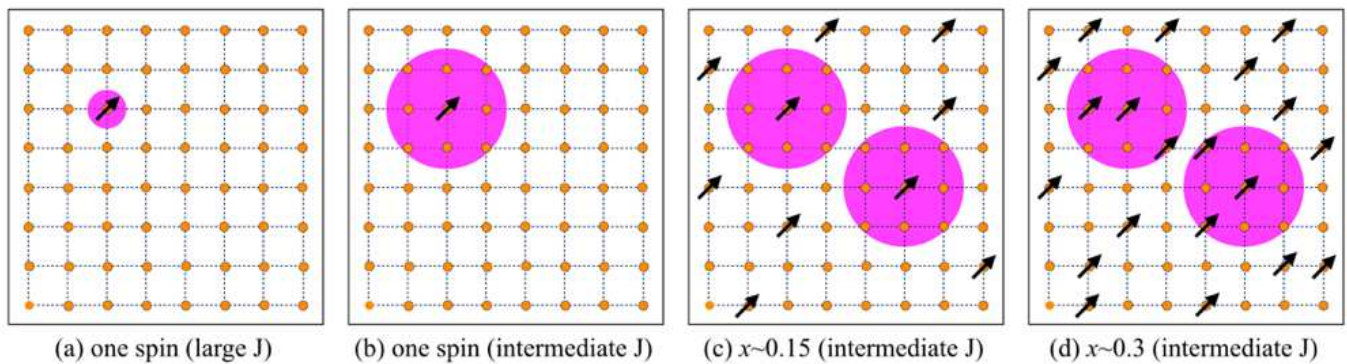}}
\vspace{.2cm}
\caption{Color online: Schematic (in 2D) shows the electron delocalization 
(polarons) by the shaded regions for four different cases; (a) localized 
polaron in the double exchange limit (large $J$) and (b) extended polaron in 
intermediate coupling regime for one spin and one electron case; extended 
polarons are shown in intermediate coupling regime for two different diluted 
limits (c) $x=0.15$ and (d) $x=0.3$.
}
\end{figure*}

Fig.3(a) shows the ferromagnetic windows for various values of $x$ in a wide 
range starting from as small as 10$\%$. We find that the optimal $p$ value where 
the $T_c$ is maximum decreases with $x$, which is in contrast to the earlier 
claim where $p$=0.5 is suggested to be the optimum value irrespective of $x$ 
values\cite{alvarez}. In experiment, both $x$ and $p$ can be changed 
simultaneously by co-doping method. It is found that $p$ decreases in the Si 
co-doped samples for all values of $x$. Consequently, for low $x$ ($\leq 0.08$) 
the $T_c$ decreases as compared to the un-codoped ones\cite{cho,wang1} and for 
higher $x$ the co-doped sample has larger $T_c$. To compare our results with 
the experiment we focus around $p=0.4$ in Fig.3(a) (the dotted line). Now, 
if we decrease $p$ the transition temperature decreases
for lower values of $x$ (=0.25 and 0.20) but increases for higher values like
$x =$ 0.30 and 0.35, which captures the experimental results discussed above.
Our calculations clearly demonstrate that the $T_{c}$ can be increased with $x$ 
provided the $p$ value is tuned properly but not arbitrarily. For $t$=0.5 eV 
the estimated $T_c$ is 120 K for $x$=0.1 which qualitatively matches with with 
the experimentally observed $T_c$ range\cite{matsukura}.

It is generally believed that the $p$ value must be maximized to obtain a higher 
ferromagnetic $T_c$ in DMSs. In Fig.3(a) our calculations show otherwise, 
that the optimum $p$ value decreases with increasing $x$. To interpret our 
finding, in Fig.3(b), we re-plot the ferromagnetic windows in terms of the absolute 
carrier density $p_{abs}$ as defined earlier. Interestingly, we find that the 
ferromagnetic windows lie on top of each other with optimum $p_{abs}$ around 
0.11 for $x$=0.35, which decreases slightly for smaller $x$ values. To understand 
this we start our discussion from the double exchange (large $J$) limit where 
carrier spins are aligned in the direction of the core spin. For $x$ = 1 
(spins at each site) carriers get delocalized and the electronic kinetic energy 
is minimized for the ferromagnetic ground state in the range 
$0$ $<$ $p_{abs}$ $<$ $1$ , where the optimum ferromagnetic $T_c$ is found to 
be at $p_{abs}$ = 0.5 [Ref. \citenum{yunoki}]. This we call the optimum $p_{abs}$. 
The range of ferromagnetic ground state is confined to $0$ $<$ $p_{abs}$ $<$ $0.3$
in the intermediate coupling regime and the optimum $p_{abs}$ decreases to 
$\sim$0.15 [Ref. \citenum{pradhan1}]. In the diluted limit ($x$=0.1-0.35) we 
find [see Fig.3(b)] that the optimum $p_{abs}$ value is $\sim$0.11 which is 
in the right ball park as compared to the $x$ = 1 limit. 
This can be understood within a polaron picture discussed below.

In the double exchange limit for one spin and one carrier problem the carrier 
remains localized to the core spin. A single site localized polaron is shown 
schematically as the shaded region in Fig.4(a). In the intermediate coupling 
regime the carrier localization extends over many lattice sites as shown in 
Fig.4(b). For a given $x$ a minimum concentration of polarons is required for 
ferromagnetic percolation. For $x$=0.15 the ferromagnetic window starts at 
$p_{abs}$ $\simeq$ 0.07 and $T_c$ is maximum for $p_{abs}$ $\simeq$ 0.10. 
When we increase $x$ the optimum $p_{abs}$ does not change in the range 
$x$=0.1-0.35, studied in this paper. This indicates that the number of spins 
in the shaded region increases without affecting the polaron size, schematically 
shown in Fig.4 (c) and (d). Now if we plot $T_{c}$ vs $x$ for $p_{abs}$ = 0.11, 
$T_{c}$ increases with $x$ in the diluted limit and saturates for concentrated 
$x$ values [see inset of Fig.3(b)]. 

Using the insight obtained from our calculations we suggest a two step procedure
to enhance the ferromagnetic $T_c$ in experiments. The first step is to determine 
the optimum carrier density $p_{abs}$ for a fixed impurity concentration $x$. So 
in this step one needs to tune $p_{abs}$ without changing $x$, which can be 
achieved by using an external process like hydrogenation\cite{theys}.
After extracting the optimal $p_{abs}$ the second step is to increase the $x$ only
without altering the optimal $p_{abs}$ value obtained in the first step. With 
increasing $x$ the $p_{abs}$ would change too, which can be tuned back to its 
optimal value by co-doping with suitable (n-type or p-type) element. 
It is important to note that co-doping not only tunes the hole density but also 
increases the effective $x$\cite{fujii,bergqvist} and will be helpful to enhance 
the $T_c$ further. We believe that a systematic combination of experimental 
processes e.g. doping, annealing, hydrogenation, and co-doping can be designed 
to prepare DMSs with higher $T_c$.  

In conclusion, our model calculations provide a new framework to increase 
the ferromagnetic $T_c$ in diluted magnetic semiconductors. The optimum 
$p_{abs}$ (absolute carrier density), where $T_c$ is maximum, lies around 0.11 
and $T_c$ increases with $x$ for fixed $p_{abs}$ in a broad range of $x$ 
studied in this letter. To replicate such a scenario in the experiment, $p_{abs}$ 
has to be determined for small $x$ and then effort should be made to increase $x$ 
without altering the $p_{abs}$ value. This procedure, viable in experiments, 
would enhance the ferromagnetic $T_c$. We hope that our 
finding will motivate new experiments by combining the growth and the 
post-growth process outlined here to achieve high $T_c$ DMSs for spintronics 
applications.

We acknowledge our discussion with Pinaki Majumdar.

\end{document}